\documentclass[12pt,twoside]{article}
\usepackage{fleqn,espcrc1}

\usepackage{graphicx,epsf}

\newcommand{\AmS}{{\protect\the\textfont2
  A\kern-.1667em\lower.5ex\hbox{M}\kern-.125emS}}

\title{Nucleon-Nucleon Scattering in a Three-Dimensional Approach}

\author{I. Fachruddin\address{Institut f\"ur Theoretische Physik II, Ruhr-Universit\"at
  Bochum, D-44780 Bochum},
        Ch. Elster\address{Insitut f\"ur Kernphysik, Forschungszentrum J\"ulich, and
     Institute for Nuclear and Particle Physics, Ohio University, Athens, OH 45701},
 W. Gl\"ockle$^a$} 
       
\begin{document}

\maketitle

\vspace{5mm}
Two-nucleon scattering at intermediate energies of a few hundred MeV requires quite a
few angular momentum states in order to achieve convergence of e.g. scattering
observables. This is even more true for the scattering of three or more nucleons upon
each other. An alternative approach to the conventional one, which is based on angular
momentum decomposition, is to work directly with momentum vectors, specifically with the
magnitudes of momenta and the angles between them.
We formulate and numerically illustrate \cite{nn3d} this alternative approach for the
case of NN scattering using two realistic interaction models, the Argonne AV18
\cite{av18} and the Bonn-B \cite{bonnb} potentials. The momentum vectors enter directly
into the scattering equation, and the total spin of the two nucleons is treated in a
helicity representation with respect to the relative momenta $\bf q$ of the two nucleons.

The momentum-helicity states are given as
\begin{equation}
\left| {\bf q};\hat{q}S\Lambda \right\rangle \equiv \left|
{\bf q}\right\rangle \left| \hat{q}S\Lambda \right\rangle = \left|
{\bf q}\right\rangle R(\hat{q})
\sum _{m_{1}m_{2}}C(\frac{1}{2}\frac{1}{2}S;m_{1}m_{2}\Lambda )\left|
\hat{z}\frac{1}{2}m_{1}\right\rangle \left| \hat{z}\frac{1}{2}m_{2}\right\rangle,
\label{eq1}
\end{equation}
where $R(\hat{q})=\exp(-iS_{z}\phi)\exp(-iS_{y}\theta)$ is the rotation operator,
$S_{z},\, S_{y}$ the components of ${\bf S}=\frac{1}{2}({\mbox {\boldmath
$\sigma$}}_1+{\mbox {\boldmath$\sigma$}}_2)$, and $\Lambda$ the eigenvalue of the
helicity operator ${\bf S}\cdot \hat{q}$. Introducing parity and two-body
isospin states $|t m_t\rangle$, the antisymmetrized two-nucleon state is given by
\begin{equation}
\left| {\bf q};\hat{q}S\Lambda ;t\right\rangle ^{\pi a}=
\frac{1}{\sqrt{2}}(1-\eta _{\pi }(-)^{S+t})\left| t\right\rangle
\left| {\bf q};\hat{q}S\Lambda \right\rangle _{\pi } \label{eq2},
\end{equation}
with the parity eigenvalues $\eta _{\pi }=\pm 1$ and 
$\left| {\bf q};\hat{q}S\Lambda \right\rangle _{\pi } \equiv  \frac{1}{\sqrt{2}} 
(\left| {\bf q} \right\rangle + \eta _{\pi} \left| -{\bf q} \right\rangle ) 
\left|\hat{q}S\Lambda \right\rangle $. 

With these basis states we formulate the Lippmann-Schwinger (LS) equations for NN
scattering. In the singlet case, $S=0$, there is one single equation for each parity,
\begin{equation}
T_{0 0 }^{\pi St}({\bf q}',{\bf q})=V_{0 0
}^{\pi St}({\bf q}',{\bf q})+\frac{1}{4} \int d^3 q''
\, V_{0 0}^{\pi St}({\bf q}',{\bf
q}'')G_{0}(q'')T_{0 0 }^{\pi St}({\bf q}'',{\bf q}) \label{eq3}
\end{equation}
Using rotational and parity invariance one finds in the triplet case, $S=1$, a set of two
coupled LS equations for each parity and each initial helicity state
\begin{eqnarray}
T_{\Lambda '\Lambda }^{\pi St}({\bf q}',{\bf q}) & = & V_{\Lambda
'\Lambda }^{\pi St}({\bf q}',{\bf q})+\frac{1}{2}\int d^3 q''\,
V_{\Lambda '1}^{\pi St}({\bf q}',{\bf q}'')G_{0}(q'')T_{1\Lambda }^{\pi
St}({\bf q}'',{\bf q})\nonumber \\
 &  & +\frac{1}{4}\int d^3 q''\, V_{\Lambda '0}^{\pi
St}({\bf q}',{\bf q}'')G_{0}(q'')T_{0\Lambda }^{\pi
St}({\bf q}'',{\bf q}).\label{eq4}
\end{eqnarray}
Because of the symmetry properties only two $ (\Lambda = 1,0) $ of the original three initial helicity states
need to be considered. Both LS equations are two-dimensional integral equations in two
variables for the half-shell t-matrix and in three variables for the fully-off-shell
t-matrix, namely two magnitudes of momenta and one angle. For explicit calculations we
choose $\hat q = \hat z$, which allows together with the properties of the potential that
the azimuthal dependence of $\bf q'$ and $\bf q$ can be factored out 
\begin{equation}
\label{eq5}
T_{\Lambda '\Lambda }^{\pi St}({\bf q}',{\bf q})=e^{i\Lambda (\phi '-\phi )}T_{\Lambda
'\Lambda }^{\pi St}(q',q,\theta ').
\end{equation}

\begin{figure}
\begin{minipage}[t]{120mm}
\centering{\resizebox*{14cm}{8cm}{\rotatebox{270}{\includegraphics{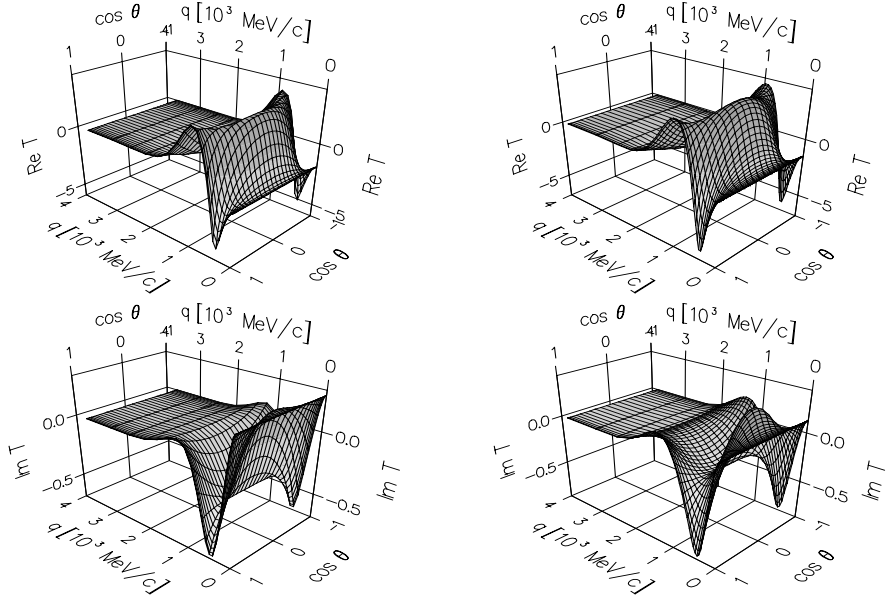}}}}
\vspace{-9mm}
\caption{$T^{101}_{00}(q,q_0,\theta)$ as a function of $q$ and
$\cos \theta$ in units of $10^{-7}$~MeV$^{-2}$. Left side for Bonn-B, right side for AV18.}
\end{minipage}
\vspace{-5mm}
\end{figure}

As it is well known, rotational-, parity-, and time reversal invariance
restricts any NN potential V to be formed out of six independent terms
\cite{wolfenstein}. We introduce the following set of linear independent
operators, which are diagonal in our basis
\begin{eqnarray}
\Omega _{1}  =  1  \: &; & \: \Omega _{2} = {\bf S}^{2}\label{40} \nonumber  \\
\Omega _{3} = {\bf S}\cdot \hat{q}'\, {\bf S}\cdot
\hat{q}' \: &; & \: \Omega _{4}={\bf S}\cdot \hat{q}'\, {\bf S}\cdot \hat{q}
\label{42} \nonumber  \\
\Omega _{5} = ({\bf S}\cdot \hat{q}')^{2}\, ({\bf S}\cdot
\hat{q})^{2} \: &; & \:
\Omega _{6} = {\bf S}\cdot \hat{q}\, {\bf S}\cdot \hat{q}\label{44}
\end{eqnarray}
and represent the most general potential in a nonrelativistic Schr\"odinger equation as linear combination of them,
\begin{equation}
\label{eq6}
 \langle {\bf q'}|V| {\bf q} \rangle \equiv
V({\bf q'},{\bf q})=\sum _{i=1}^{6}v_{i}(q',q,\gamma) \Omega_{i}.
\end{equation}
Here $v_{i}(q',q,\gamma)$ are scalar functions of the magnitudes of the vectors 
$\bf q'$,  $\bf q$, and the angle $\gamma={\hat q}'\cdot {\hat q}$ between them. It
should be noted that for $\bf q$ pointing in $z$-direction one can use 
$\left\langle \hat{q}'S\Lambda '\right. \left| \hat{z}S\Lambda \right\rangle =\exp(i\Lambda
(\phi '-\phi ))d^{S}_{\Lambda \Lambda '}(\theta ')$, so that the azimuthal dependence
factors out of V and thus also out of the T-matrix. The operators $\Omega_{i}$ can be
transformed to the standard Wolfenstein form \cite{wolfenstein} with a linear
transformation as shown in \cite{nn3d}.


To demonstrate the feasibility of our formulation when applied to NN scattering we
present numerical results for two different NN potentials of quite different character,
the Bonn-B \cite{bonnb} and the AV18 \cite{av18} potential models. Similarities and
differences in the half-shell T-matrices $T^{\pi St}_{\Lambda' \Lambda}(q,q_0,\theta)$
are displayed for $T^{101}_{00}(q,q_0,\theta)$ for $q_0=375~MeV/c$ in Fig.~1. 
The amplitudes are identical on-shell and show deviations off-shell. 
The choice presented
in Fig.~1 is quite representative for the off-shell differences between the two models.

\begin{figure}
\begin{minipage}[t]{120mm}
\resizebox*{6cm}{5.5cm}{\rotatebox{270}{\includegraphics{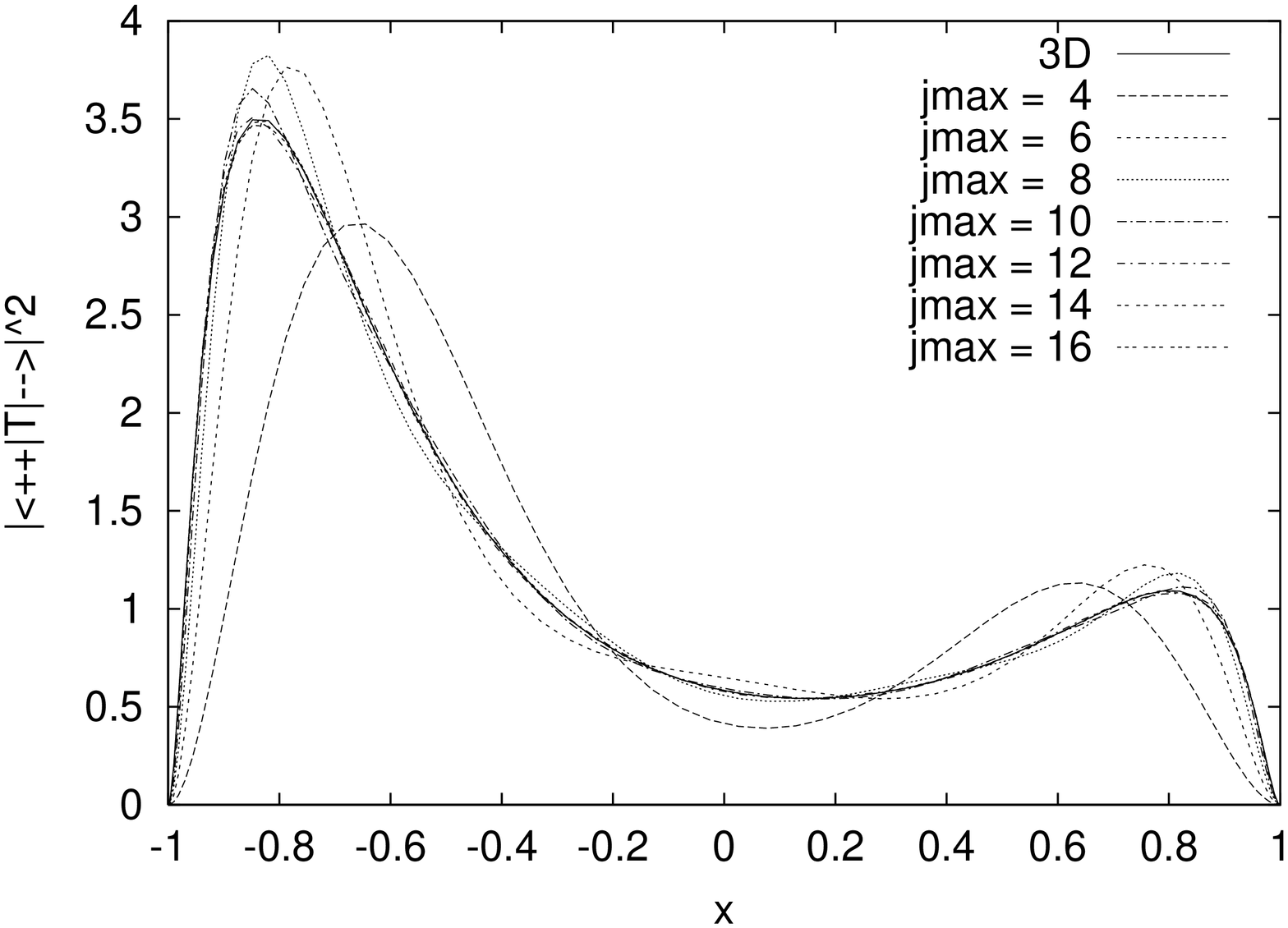}}} 
\resizebox*{6cm}{5.5cm}{\rotatebox{270}{\includegraphics{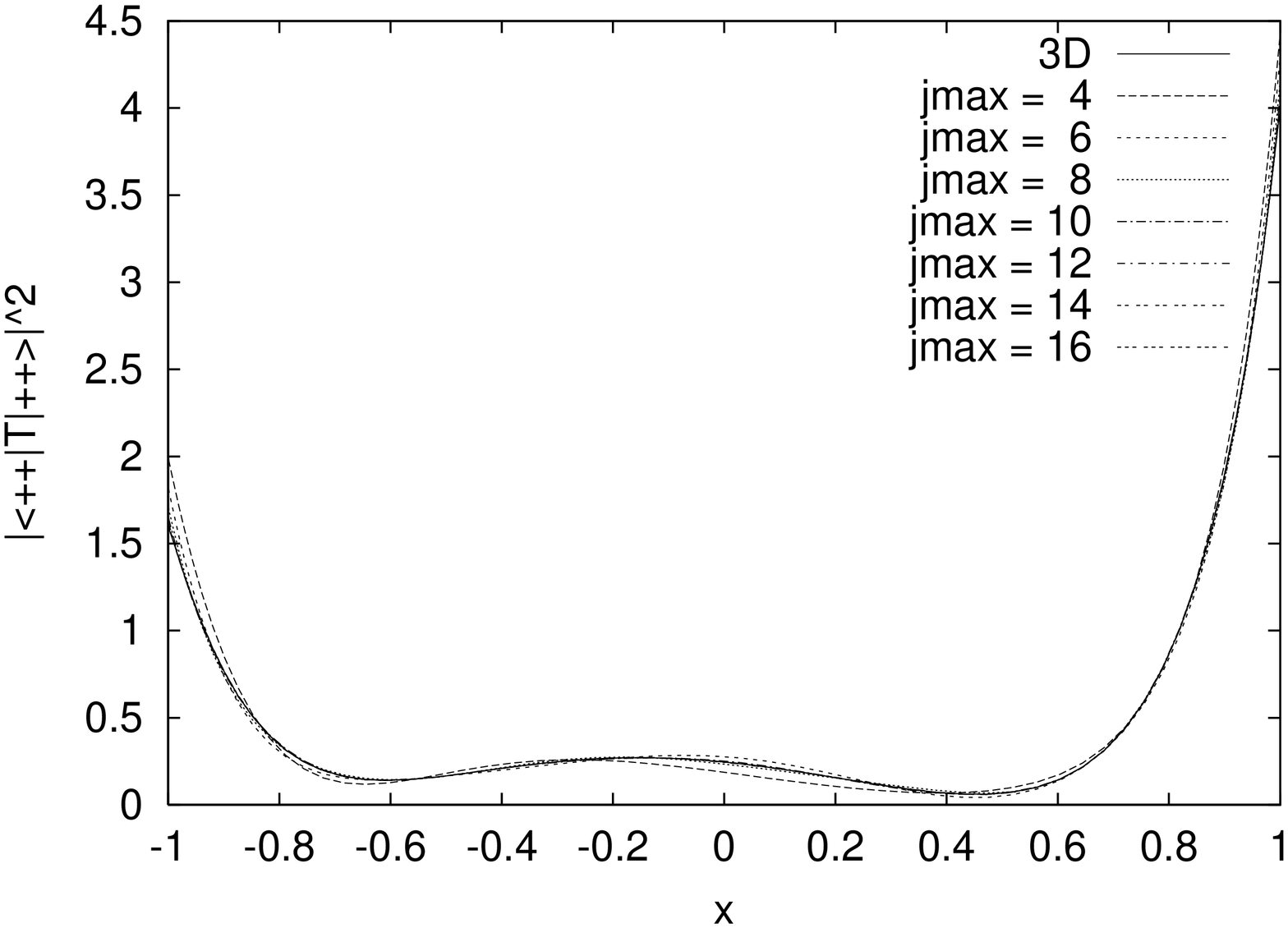}}}
\vspace{-6mm}
\caption{$|\langle ++|T|--\rangle|^2$ and $|\langle ++|T|++\rangle|^2$ in units of 
$10^{-14}$~MeV$^{-4}$ as a function of $x=\cos \theta$ for $q_0$=375~MeV/c.}
\end{minipage}
\vspace{-5mm}
\end{figure}

In order to connect with standard representations of the NN t-matrix as well as to
calculate observables, we express $T$ in terms of antisymmetric states
\begin{equation}
\label{22}
\left| \tau _{1}\tau _{2}m_{1}m_{2}{\bf q}\right\rangle _{a}\equiv
\frac{1}{\sqrt{2}}(1-P_{12})\left| \tau _{1}\tau
_{2}m_{1}m_{2}{\bf q}\right\rangle ,
\end{equation}
where $\tau_i$ and $ m_i$ are the magnetic isospin and spin quantum numbers and $P_{12}$
is the permutation operator of the two nucleons. 
In this representation, which we denote as `physical', we
obtain for the on-shell T-matrix elements ( $|\bf q'|=|\bf q|$)
\begin{eqnarray}
_{a}\langle \tau _{1}\tau
_{2}m_{1}'m_{2}'q\hat{q}'|& T &| \tau _{1}\tau _{2}m_{1}m_{2}{\bf q}\rangle
_{a}=  
 \frac{1}{4}e^{-i(\Lambda _{0}'-\Lambda _{0})\phi '}\sum _{S\pi
t}C(\frac{1}{2}\frac{1}{2}t;\tau _{1}\tau _{2})^{2}(1-\eta _{\pi }(-)^{S+t})\nonumber \\
 &  & \quad C(\frac{1}{2}\frac{1}{2}S;m_{1}'m_{2}'\Lambda
_{0}')C(\frac{1}{2}\frac{1}{2}S;m_{1}m_{2}\Lambda _{0})\sum _{\Lambda '}d^{S}_{\Lambda
_{0}'\Lambda '}(\theta ')T_{\Lambda '\Lambda _{0}}^{\pi St}(q,q,\theta ').\label{28}
\end{eqnarray}
The quantity $T_{\Lambda '\Lambda _{0}}^{\pi St}(q,q,\theta ')$ can be related through 
straightforward but lengthy algebra 
to the standard partial wave T-matrix $T^{Sjt}_{l'l}(q)$ as is shown in \cite{nn3d}.
Here we want to consider the on-the-energy-shell amplitudes $|_a\langle \tau_1 \tau_2 m_1'
m_2' q_0 \hat{q}|T|\tau_1 \tau_2 m_1 m_2 {\bf q_0}\rangle
_a|^2 \equiv  |\langle m_1' m_2' | T| m_1 m_2 \rangle |^2$, $m=\pm \frac{1}{2}$. 
If one takes rotational symmetry and parity invariance
into account, one ends up with six independent amplitudes out of the 16 possible
combinations. Two of them, $|\langle ++|T|--\rangle|^2$ and $|\langle ++|T|++\rangle|^2$,
are displayed in Fig.~2 as a function of $\cos \theta$ where $\theta$ is c.m. angle for a projectile energy of
300~MeV ($q_0$=375~MeV/c). In addition to the result from the 3D calculation we show the partial wave
sums for increasing $j_{max}$. One can clearly see that at $E_{lab}$=300~MeV partial
waves up to at least $j_{max}$=12 are needed to obtain a converged result for every matrix element.

From the `physical' T-matrix amplitudes we construct the Wolfenstein 
amplitudes \cite{Glo83}. Once those are obtained it is straightforward to 
calculated the NN scattering observables \cite{hoshizaki}. In order to unambiguously
define the normalization we also give the expression
for the spin-averaged differential cross section for nucleon species
$\tau_{1} \tau_{2}$ as
\begin{equation}
\frac{d\sigma}{d\Omega} = (2\pi)^4 \left(\frac{m}{2}\right)^2 \frac{1}{4}
\sum_{m'_1 m'_2 m_1 m_2}  |_a\langle \tau_{1} \tau_{2} m_{1}' m_{2}' q {\hat q}'
 | T | \tau_{1} \tau_{2} m_{1} m_{2} {\bf q} \rangle_a |^2
\label{eq:5.13}
\end{equation}
In Fig.~3 the differential cross section  is given for
$E_{lab}$=300~MeV together with partial wave sums for increasing $j_{max}$.
It should be noted that the convergence of the partial wave sums is especially
slow for the backward angle differential cross section. A sum up to $j_{max}$=16 is
needed to obtain convergence within 1 \%. In Fig.~4 $A_y$ is displayed, which needs a partial
wave sum up to $j_{max}$=10 to reach a similar convergence.

\begin{figure}[htb]
\begin{minipage}[t]{75mm}
\centering{\resizebox*{6.5cm}{4cm}{\rotatebox{270}{\includegraphics{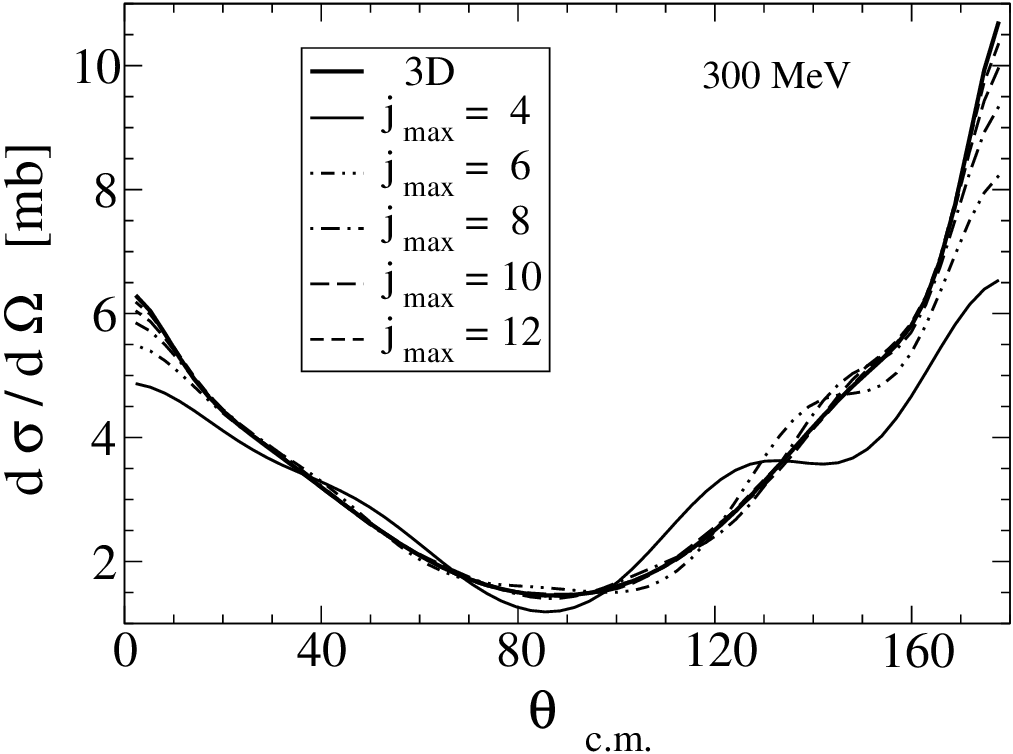}}}}
\vspace{-6mm}
\caption{$\frac {d \sigma}{d \Omega}$ as a function of $\theta$ for
$q_0$=375~MeV/c.}
\end{minipage}
\hspace{\fill}
\begin{minipage}[t]{75mm}
\centering{\resizebox*{6.5cm}{4cm}{\rotatebox{270}{\includegraphics{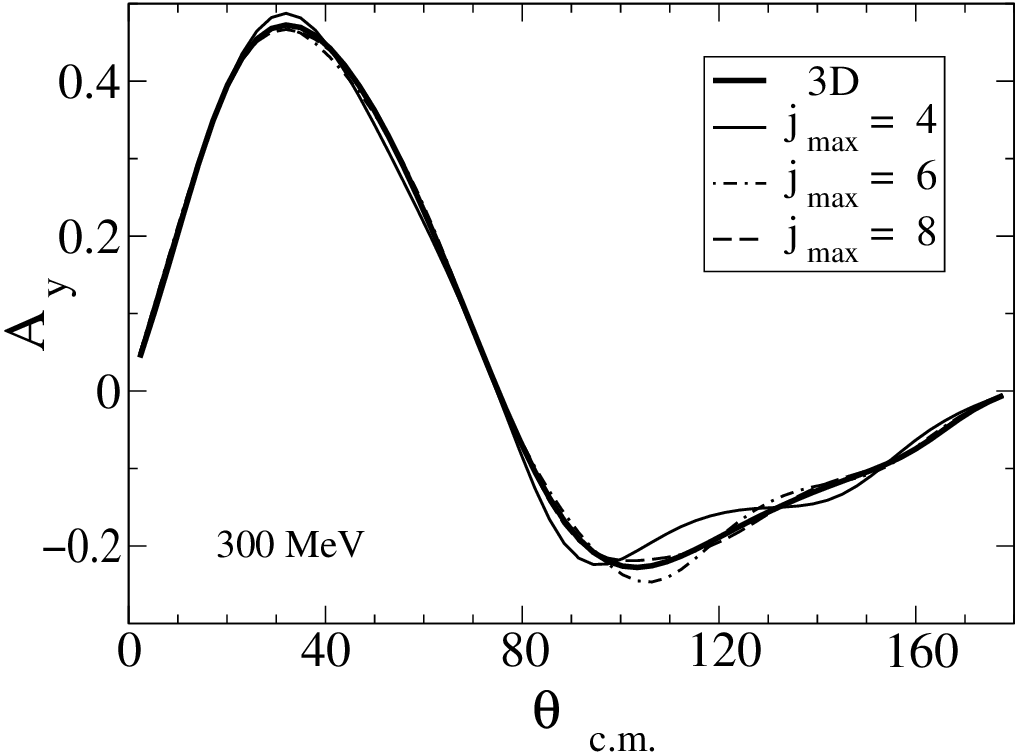}}}}
\vspace{-6mm}
\caption{$A_y$ as a function of $\theta$ for $q_0$=375~MeV/c.}
\end{minipage}
\end{figure}

\vspace{-5mm}
In summary, we formulated and numerically illustrated an alternative approach to NN
scattering which works directly with momentum vectors. The spin of the two nucleons is
treated in a helicity representation with respect to the relative momentum of the two
nucleons. 
We would like to emphasize that the here developed scheme is
algebraically quite simple to handle provided potentials are given in an operator form.
This is e.g. the case for all interactions developed within a field theoretic frame work.
This work is intended to serve as starting point towards treating three-nucleon
scattering in a similar fashion.

\end{document}